# HAWC J2227+610: a potential PeVatron candidate for the CTA in the northern hemisphere

**Gaia Verna,[a,*] Franca Cassol[a] and Heide Costantini[a] on behalf of the CTA Consortium**
(a complete list of authors can be found at the end of the proceedings)

[a] *Aix Marseille Univ, CNRS/IN2P3, CPPM, Marseille, France*
*E-mail:* verna@cppm.in2p3.fr, cassol@cppm.in2p3.fr, costant@cppm.in2p3.fr

Recent observations of the gamma-ray source HAWC J2227+610 by Tibet AS+MD and LHAASO confirm the special interest of this source as a galactic PeVatron candidate in the northern hemisphere. HAWC J2227+610 emits Very High Energy (VHE) gamma-rays up to 500 TeV, from a region coincident with molecular clouds and significantly displaced from the nearby pulsar J2229+6114. Even if this morphology favours an hadronic origin, both leptonic or hadronic models can describe the current VHE gamma-ray emission. The morphology of the source is not well constrained by the present measurements and a better characterisation would greatly help the understanding of the underlying particle acceleration mechanisms. The Cherenkov Telescope Array (CTA) will be the future most sensitive Imaging Atmospheric Cherenkov Telescope and, thanks to its unprecedented angular resolution, could contribute to better constrain the nature of this source. The present work investigates the potentiality of CTA to study the morphology and the spectrum of HAWC J2227+610. For this aim, the source is simulated assuming the hadronic model proposed by the Tibet AS+MD collaboration, recently fitted on multi-wavelength data, and two spatial templates associated to the source nearby molecular clouds. Different CTA layouts and observation times are considered. A 3D map based analysis shows that CTA is able to significantly detect the extension of the source and to attribute higher detection significance to the simulated molecular cloud template compared to the alternative one. CTA data does not allow to disentangle the hadronic and the leptonic emission models. However, it permits to correctly reproduce the simulated parent proton spectrum characterized by a ~ 500 TeV cutoff.

*Presenter

## 1. Introduction

The search for galactic PeVatrons, astrophysical particle sources capable of accelerating cosmic rays (CR) up to PeV energies, is an ambitious challenge of the multi-wavelength and the multi-messenger astronomy of this century. In the field of gamma-ray astronomy several experiments are currently contributing to detect galactic gamma-ray sources significantly above hundreds of TeV. These sources indicate the presence of highly energetic charged particles, which may suggest the presence of a PeVatron accelerator nearby. In 2020, the HAWC observatory has detected a significant emission above 100 TeV in the direction of VER J2227+608 [1], associated with the supernova remnant (SNR) G106.3+2.7 [2]. Other recent observations performed by Tibet AS+MD [3] and by the LHAASO [4] experiments have highlighted the particular interest of this source from which gamma photons up to 500 TeV were detected. Not only ground based experiments, but also the Fermi-LAT satellite [5] had earlier significantly detected this source in the 3-500 GeV energy range. The observed gamma-ray emission had been proven to be spatially extended ($\sim 0.2°$) and interestingly correlated with the molecular clouds in the SNR region. This morphological feature, together with a significant angular displacement ($\sim 0.4°$) of the Very High Energy (VHE) emission region from the nearby pulsar PSR J2229+6114, appears to favour a hadronic origin of the observed gamma-radiation. Nevertheless, the current VHE data allow for both hadronic or leptonic interpretation of the underlying emission mechanism. Similarly the morphology of the source is not well constrained due to the not optimal angular resolution of the current gamma-ray instruments.

In this work, we investigate the potential of the future Cherenkov Telescope Array (CTA) [6] for the morphological and spectral characterization of HAWC J2227+610. This source will only be observable from the CTA-North site, located in La Palma (Canary Islands). The array is currently under construction and it is foreseen to reach a first construction configuration (*alpha*) and than to be enlarged in a full-scope configuration (*omega*) characterized by a greater number of telescopes. It is worth mentioning that in the *omega* configuration, the CTA-North array is expected to have more than four times better sensitivity (at 1 TeV) compared to VERITAS and a factor two better angular resolution ($\sim 0.04°$ at 10 TeV).

Future observations of HAWC J2227+610 with the CTA-North array have been simulated considering the most recent fitted models of the VHE emission. The study is divided into two parts, the first centered on the morphological characterization of the source (Sec. 3.1) and the second on its spectral modeling (Sec. 3.2). Conclusions and perspectives are presented in Sec. 4.

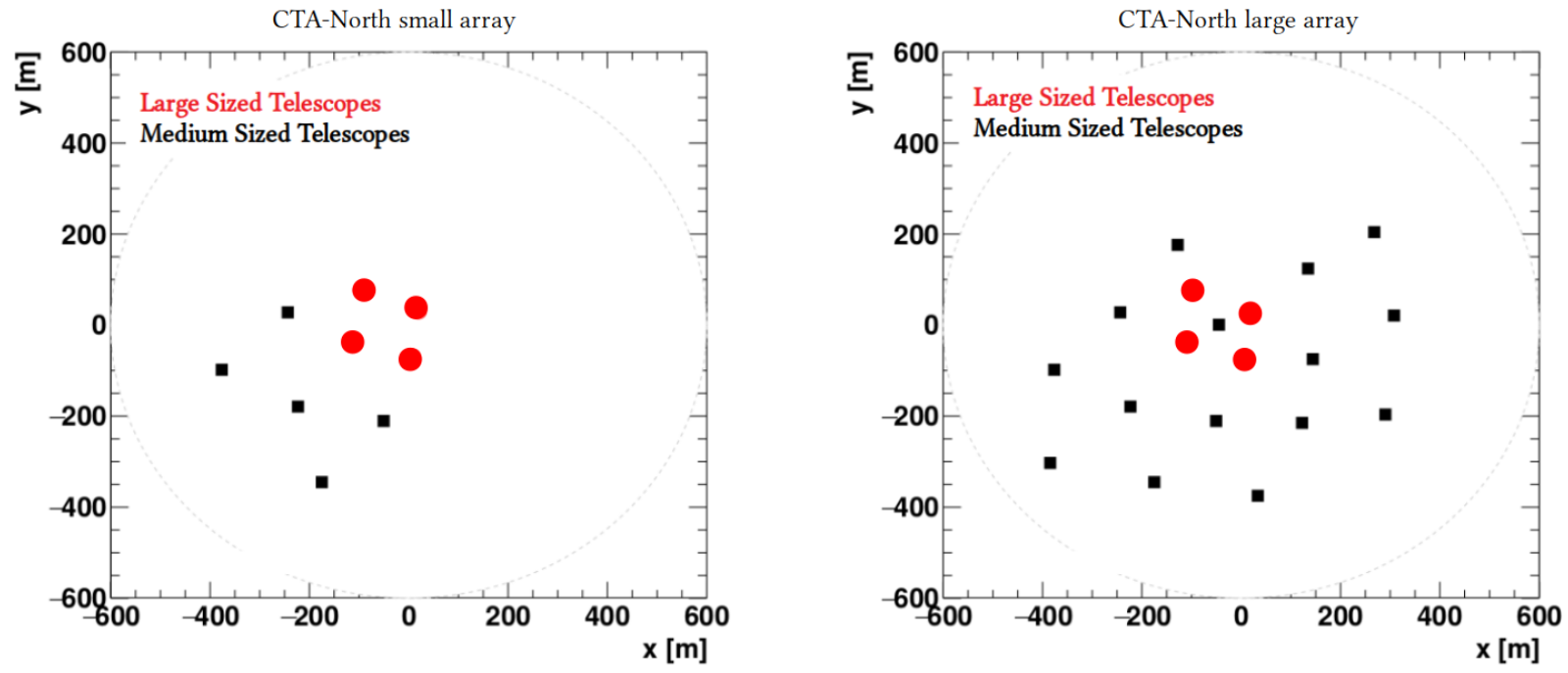

**Figure 1:** CTA-North layouts considered in this work: *small* (left) and *large* (right) configurations.

## 2. Simulations and analysis

The present study is based on the future CTA science tool, `gammapy` (v18.2) [7], which allows to build a full analysis workflow, starting from the simulation of synthetic data to the production of high level scientific results such as sky-maps, spectra and flux points.

The source is simulated in a $2° \times 2°$ field of view, centered on the position of VER J2227+608, with a spatial bin size of 0.01°. The emission is studied in a 30 GeV - 160 TeV energy window, 10 bins per decade. For simulation and analysis the *prod3b* CTA Instrument Response Functions (IRFs) have been used [8] in which a *small* configuration corresponds to 9 telescopes (4 Large Sized Telescopes and 5 Medium Sized Telescopes with mirrors of 23 m and 12 m diameter respectively) and a *large* configuration amounts to 19 telescopes (4 Large and 15 Medium-Sized ), as shown in Fig. 1. These configurations do not correspond to the final CTA design, which is still in a finalisation phase. IRFs include the effective area, the angular and energy resolution, and the background rate from cosmic rays mis-reconstructed as gammas. HAWC J2227+610, observed from the La Palma site location, culminates at almost 50°/60° altitude above the horizon, therefore we consider IRFs with 40° zenith angle pointing direction. In addition to the proton mis-reconstructed background, we include a background from diffuse gamma-rays adopting the model used for the CTA Galactic Plane Survey [9]. We assume that the observed gamma-radiation from HAWC J2227+610 is produced by the interaction of relativistic protons with the molecular clouds in the neighbourhood of the source, via the pion decay channel. The hadronic modeling is performed with `Pythia8` parametrization provided by the package `naima` (v0.9.1) [10] that computes non-thermal radiation from relativistic particle populations. We adopt the model fitted by the Tibet AS+MD on multi-wavelength data in the energy range 3 GeV - 114 TeV [3]. They obtained a power law proton distribution with a spectral index of 1.79 and an exponential cutoff at 499 TeV. The fitted total proton energy above 1 GeV is $W_p = 5 \times 10^{47}$ erg for a target density of 10 cm$^{-3}$. They assumed a source distance of 800 pc. The morphology of the source is modeled using the radio template map of molecular hydrogen in the direction of HAWC J2227+610, obtained through the $^{12}$CO(J=1-0) emission lines. The radio data considered in this work have been obtained from the FCRAO survey [11]. We consider the two maps associated to the Fermi gamma-emission and the VERITAS gamma-emission, which are obtained integrating the radio cube in two different doppler velocity windows: $[-4, -6]$ km/s and $[-5.59, -7.23]$ km/s, respectively. The source spatial templates are obtained selecting map regions with density above $2 \times 10^{20}$ cm$^{-2}$, from which we assume the gamma emission arises. The resulting source shape is the area of the maps contained inside the black contours. In the following, we will refer to the two shapes as Template A (Fig. 2-left) and Template B (Fig. 2-right).

Three observation times, 50, 100 and 200 hours, are taken into account. In each configuration we produce a Monte Carlo sample of 100 simulations which is then analysed in order to statistically characterize the morphological and spectral features of the observed source.

We perform a 3D map based analysis. The simulated counts are stored in data cubes with longitude and latitude coordinates along two axes and energies along the third dimension. We test several model hypothesis performing 3D fits that take into account the spectral and morphological model at the same time. The fit is performed with the `sherpa` (v4.12.0) backend and the `simplex` optimization algorithm [12]. The hypothesis test is based on the cstat statistics [13] defined as $C = 2 \times \log L$, where $L$ is the fit likelihood. The change in cstat from one fitted model to another, $\Delta C$, is used for statistical test of alternative hypothesis. To be able to interpret the results in this sense, after the fit we evaluate the cstat in re-binned data cubes in order to increase the count statistics per bin. Thus, for the comparison between the various morphological models we integrate the counts along the energy axis leaving the spatial matrix unchanged. Dealing with the spectral characterization, instead, we sum up the counts on the two spatial axes, maintaining a maximum number of 18 energy bins between 30 GeV and 160 TeV. The results obtained from the morphological and spectral study of HAWC J2227+610 are described in Sec. 3.

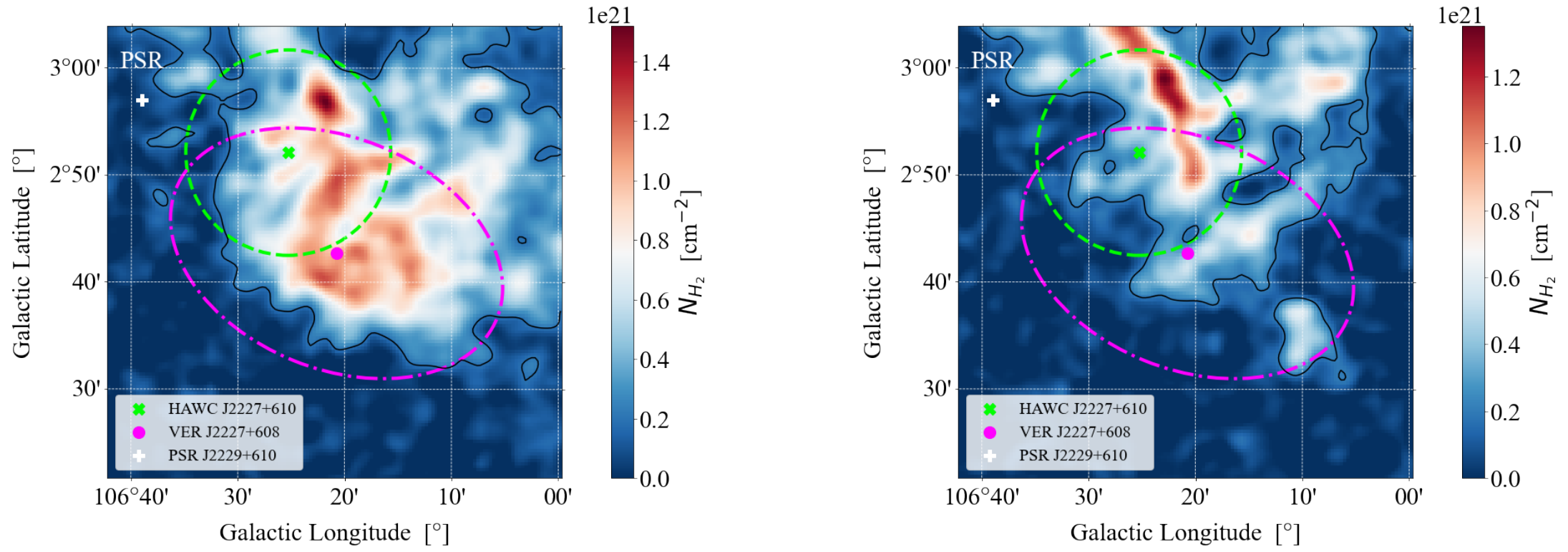

**Figure 2:** Column density maps from FCRAO survey [11] obtained integrating the measured brightness temperature over two different doppler velocity windows: $[-4, -6]$ km/s (left) and $[-5.59, -7.23]$ km/s (right). The dashed and dashed-dotted lines encompass the gamma-ray emission detected by HAWC and VERITAS, respectively. The HAWC circle marks the uncertainty of the fitted position whereas the ellipse represents the best fitted elongated shape found by VERITAS. The position of the nearby pulsar is also shown as a white cross. The black contours mark the $2 \cdot 10^{20}$ cm$^{-2}$ column density level which enclose the template area used for modeling the source. They are referred in the text as Template A (left) and Template B (right).

## 3. Results

### 3.1 Morphological Study

As described in Sec. 2, the hadronic emission is assumed to be spatially coincident with the position of the molecular clouds and simulated with the templates A and B. The data cubes are then fitted with the following spatial model: templates A and B, point source, symmetric and asymmetric gaussian and disk. For each model $\Delta C$ is computed with respect to a null hypothesis which includes only background data (mis-reconstructed protons and diffuse gammas). Fig. 3 shows the average $\Delta C$ of 100 realizations for both templates, the point source and the symmetric gaussian alternative hypothesis, as a function of the observation time and the *small* and the *large* array. As expected, the highest test statistic is obtained, in all cases, using as alternative hypothesis the simulated template, while the lowest $\Delta C$ is obtained with a point source hypothesis. The point source hypothesis is indeed excluded, with a $5\sigma$ confidence level, if compared, as null hypothesis, to the (nested) gaussian or disk models. Thus, the extension of the source is clearly detected. The gaussian and disk models, both symmetric and asymmetric, achieve equivalent detection significance with respect to the simulated molecular cloud template (in Fig. 3 we show only results from the symmetric gaussian model), while the alternative molecular cloud template reaches significantly lower $\Delta C$ values. Hence, the spatial fit does not permit to identify without ambiguity the original source morphology, but it permits to attribute higher detection significance to the simulated molecular cloud template, disfavouring the alternative one. Fig. 4 presents one example of simulated excess map and its best fitted symmetric gaussian extension, obtained with the *small* or *large* arrays, after 200 hours of observation. The averaged fitted standard deviation for the gaussian model is $\sigma = 0.145° \pm 0.001°$ and the disk radius is $r = 0.243° \pm 0.002°$, with the large array.

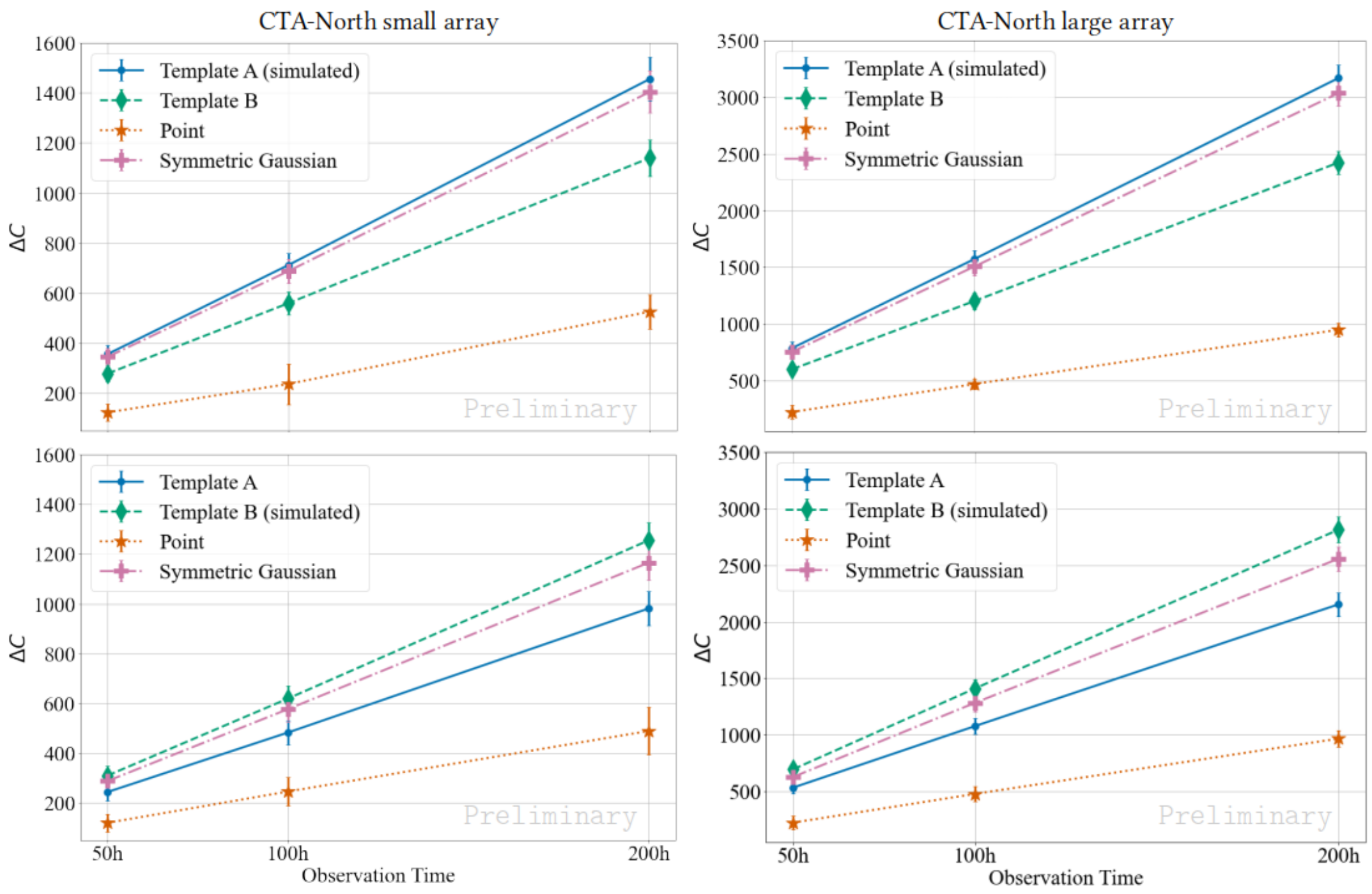

**Figure 3:** Mean values of $\Delta C$ from 100 realisations as a function of the observation time. Error bars mark the standard deviations of each distribution. Results are shown for the *small* (left) and *large* (right) array, and for the simulated template A (top) and the simulated template B (bottom).

### 3.2 Spectral Study

The spectral study is performed simulating only the Template A spatial model and, as previously, assuming the hadronic spectral model proposed by Tibet AS+MD. As for the morphology study, we fit the observed data considering different alternative hypothesis: an hadronic model, a leptonic model and three simple gamma-ray emission hypothesis that do not involve underlying physical assumptions for the radiative mechanism: a power law (PL), a power law exponential cutoff (PLEC) and a LogParabola. The leptonic fit is performed with the model defined by Tibet AS+MD, which assumes a PLEC energy distribution for the parent electrons and their interaction with the CMB and infrared photon fields.

The parameter distributions obtained with the hadronic fit of the 100 simulations are centred on the true simulated parameters with the relative dispersion shown in Tab. 1 for all parameters and simulated configurations. The precision of the fit is increasing of about a factor two from 50 to 200 hours of observation time while the larger layout permits to improve the precision of less then a factor two. Similarly, the leptonic fit permits to recover the Tibet AS+MD leptonic fit parameters, which reflects the degeneration between the two models. In fact, Fig. 5 shows that the average test statistic $\Delta C$ obtained with the different alternative hypothesis is equivalent for all hypothesis, except for the PL model, which achieves slightly lower average $\Delta C$ above 50 hours of observation. Indeed, if we quantify the significance of the presence of a spectral curvature (cutoff) considering the (nested) PLEC and LogParabola models as alternative hypothesis to the PL model (null hypothesis), the alternative hypothesis proves to be preferable, at more than $5\sigma$ confidence level, for almost all configurations and observation times, as shown in Fig. 6. The cutoff fitted with PLEC is around 50 TeV. Finally, Fig. 7 presents an example of flux points, as measured by the *small* and *large* arrays for 50 hours of observation, together with the spectral points measured by the other experiments.

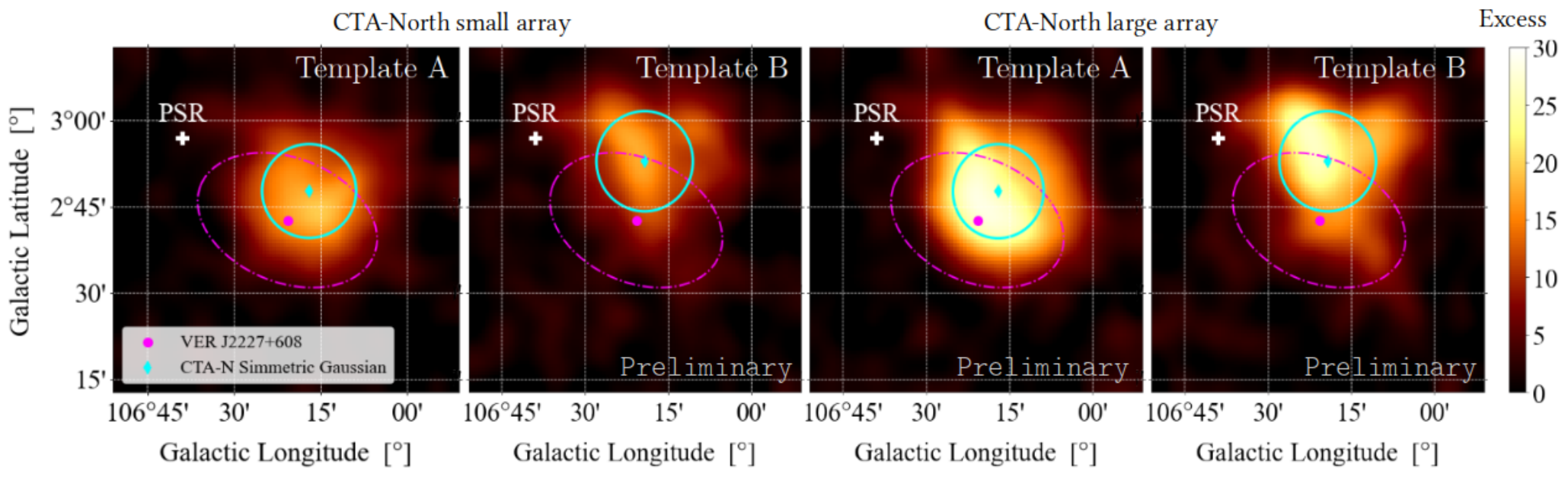

**Figure 4:** 200 hours simulated excess maps of HAWC J2227+610 observed by CTA-North in the *small* (**left**) and *large* (right) configurations. The maps are smoothed with a 0.05° gaussian sigma corresponding to the CTA angular resolution at 1 TeV. The employed spatial template is indicated in each figure. The solid circle marks the best fit of the source extension (standard deviation) assuming a symmetric gaussian model. The dashed-dotted ellipse encompasses the gamma-ray emission detected by VERITAS. The position of the nearby pulsar is also shown as a white cross.

| | amplitude | | spectral index | | energy cutoff | |
|---|---|---|---|---|---|---|
| | *small* | *large* | *small* | *large* | *small* | *large* |
| 50 h | 4.4% | 3.0% | 2.4% | 2.0% | 36.0% | 20.7% |
| 100 h | 2.9% | 2.3% | 1.6% | 1.5% | 20.1% | 15.2% |
| 200 h | 2.0% | 1.6% | 1.3% | 0.8% | 14.2% | 10.6% |

**Table 1:** Relative dispersion (standard deviation over mean) of the fitted parameters in the case of the hadronic model, as a function of the array configurations (*small*, *large*) and observation times (50, 100 and 200 hours).

## 4. Conclusions and outlook

In the present work we have investigated the CTA potentiality in constraining the morphology and the spectrum of HAWC J2227+610, a source recently detected at energies above 100 TeV by several experiments. For its spectral and spatial characteristics, HAWC J2227+610 is a strong PeVatron candidate, therefore a particularly interesting target for the future CTA. The simulations

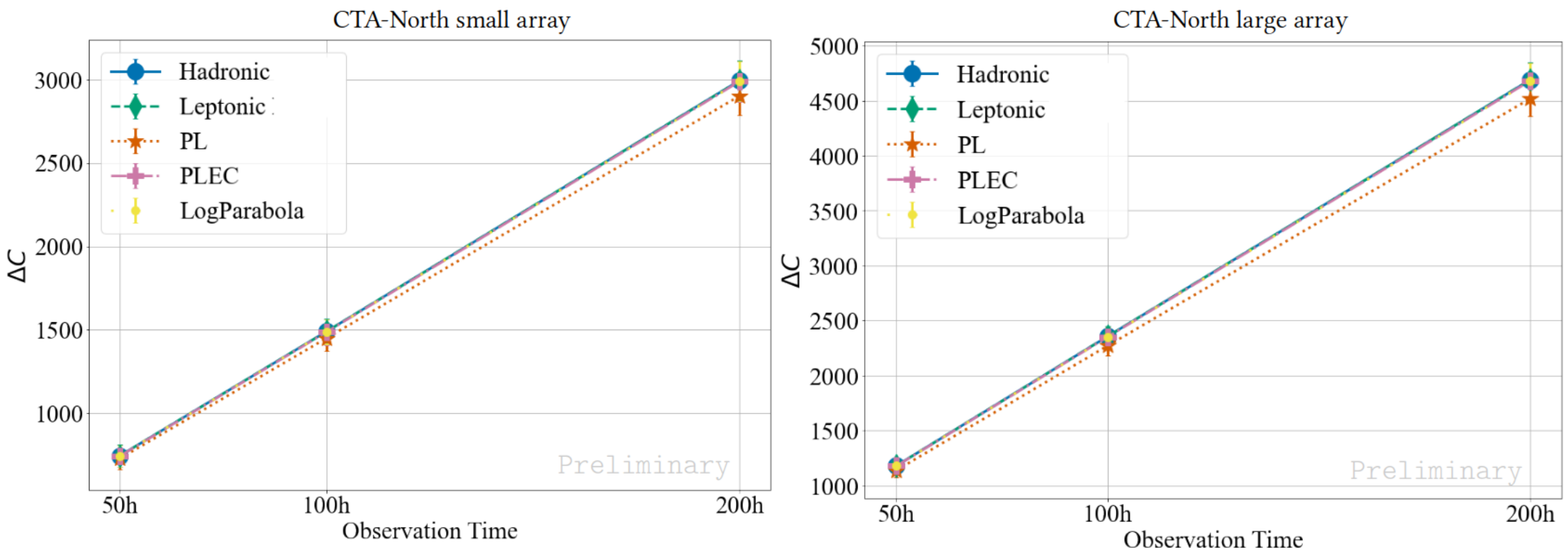

**Figure 5:** Average $\Delta C$ of 100 realizations for the hadronic, leptonic, power law (PL), power law exponential cutoff (PLEC) and LogParabola alternative hypothesis, as a function of the observation time for the *small* and *large* array.

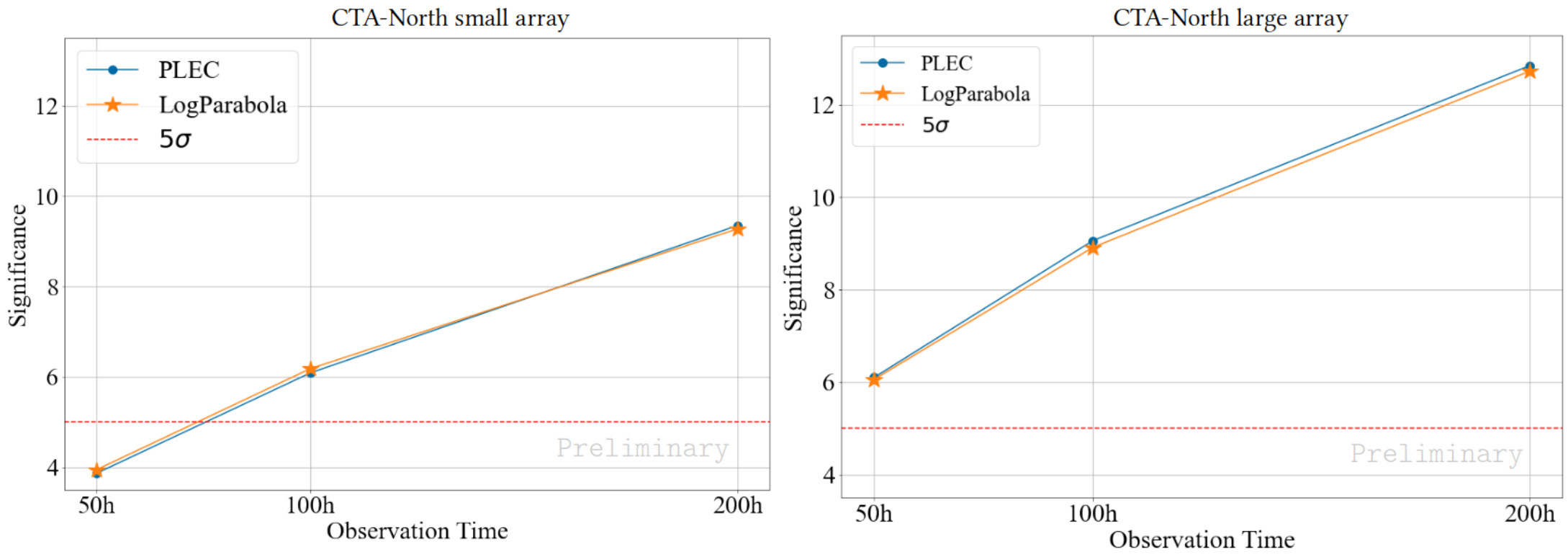

**Figure 6:** Significance (expressed in number of $\sigma$) of a spectral cutoff as a function of the observation time. The alternative power law exponential cutoff (PLEC) and LogParabola hypothesis are tested versus the power law (PL) model. The significance is evaluated assuming a $\chi^2$ distribution for the test statistic with 1 degrees of freedom for the PLEC or 2 degrees of freedom for PLEC and LogParabola hypothesis.

and the analysis were performed using the CTA science tool, `gammapy`, assuming an hadronic emission model and two spatial models, extracted from the source nearby molecular cloud maps. Two layouts for the CTA-North array (*small* and *large*) and different observations times (50, 100 and 200 hours) have been considered. In all these configurations, the source is significantly detected over the background (well above a $5\sigma$ confidence level). The extended spatial hypothesis is significantly preferred over a simple point source hypothesis. The fitted standard deviation and radius, in the case of a symmetric gaussian and disk model, are $\sim 0.14°$ and $\sim 0.24°$, respectively. The simulated molecular cloud models obtain detection significance similar to simple parametric models (gaussian and disk). However, they achieve significantly higher test statistic values with respect to the alternative molecular cloud model, which is encouraging for future morphological studies. Regarding the spectral study, CTA-North is not able to disentangle the hadronic emission assumed in this work from a leptonic one. However, it is able to correctly reproduce the parameters of the simulated hadronic model and to detect a ~50 TeV exponential cutoff in the gamma-ray

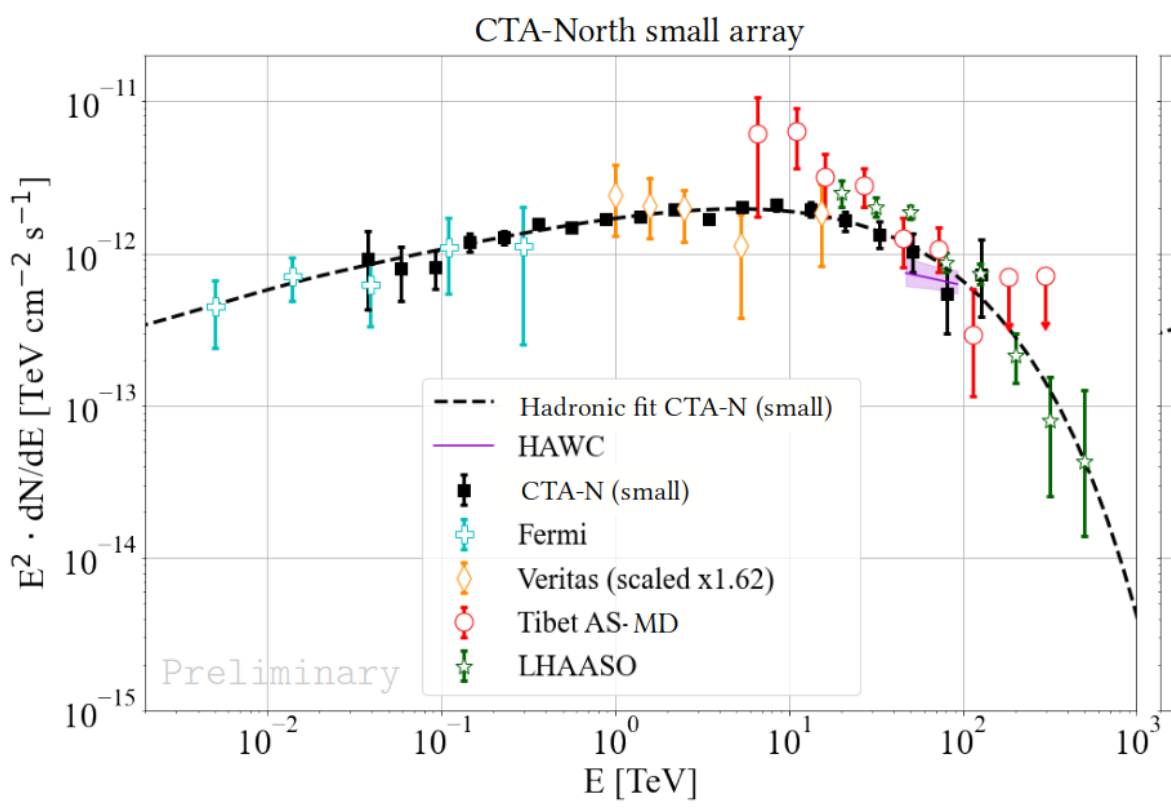

**Figure 7:** Flux points of HAWC J2227+610 as seen by CTA-North in the *small* and *large* configurations after 50 hours of observation. They are obtained assuming the best fitted hadronic model which is shown as a dashed black line. The flux points from Fermi-LAT, VERITAS, Tibet AS+MD and LHAASO are also shown.

spectrum with more than $3\sigma$ confidence level.

## Acknowledgments

We gratefully acknowledge financial support from the agencies and organizations listed here: http://www.cta-observatory.org/consortium_acknowledgments

**The Cherenkov Telescope Array Consortium July 2021 Authors**

H. Abdalla[1], H. Abe[2], S. Abe[2], A. Abusleme[3], F. Acero[4], A. Acharyya[5], V. Acín Portella[6], K. Ackley[7], R. Adam[8], C. Adams[9], S.S. Adhikari[10], I. Aguado-Ruesga[11], I. Agudo[12], R. Aguilera[13], A. Aguirre-Santaella[14], F. Aharonian[15], A. Alberdi[12], R. Alfaro[16], J. Alfaro[3], C. Alispach[17], R. Aloisio[18], R. Alves Batista[19], J.-P. Amans[20], L. Amati[21], E. Amato[22], L. Ambrogi[18], G. Ambrosi[23], M. Ambrosio[24], R. Ammendola[25], J. Anderson[26], M. Anduze[8], E.O. Angüner[27], L.A. Antonelli[28], V. Antonuccio[29], P. Antoranz[30], R. Anutarawiramkul[31], J. Aragunde Gutierrez[32], C. Aramo[24], A. Araudo[33,34], M. Araya[35], A. Arbet-Engels[36], C. Arcaro[1], V. Arendt[37], C. Armand[38], T. Armstrong[27], F. Arqueros[11], L. Arrabito[39], B. Arsioli[40], M. Artero[41], K. Asano[2], Y. Ascasíbar[14], J. Aschersleben[42], M. Ashley[43], P. Attinà[44], P. Aubert[45], C. B. Singh[19], D. Baack[46], A. Babic[47], M. Backes[48], V. Baena[13], S. Bajtlik[49], A. Baktash[50], C. Balazs[7], M. Balbo[38], O. Ballester[41], J. Ballet[4], B. Balmaverde[44], A. Bamba[51], R. Bandiera[22], A. Baquero Larriva[11], P. Barai[19], C. Barbier[45], V. Barbosa Martins[52], M. Barcelo[53], M. Barkov[54], M. Barnard[1], L. Baroncelli[21], U. Barres de Almeida[40], J.A. Barrio[11], D. Bastieri[55], P.I. Batista[52], I. Batkovic[55], C. Bauer[53], R. Bautista-González[56], J. Baxter[2], U. Becciani[29], J. Becerra González[32], Y. Becherini[57], G. Beck[58], J. Becker Tjus[59], W. Bednarek[60], A. Belfiore[61], L. Bellizzi[62], R. Belmont[4], W. Benbow[63], D. Berge[52], E. Bernardini[52], M.I. Bernardos[55], K. Bernlöhr[53], A. Berti[64], M. Berton[65], B. Bertucci[23], V. Beshley[66], N. Bhatt[67], S. Bhattacharyya[67], W. Bhattacharyya[52], S. Bhattacharyya[68], B. Bi[69], G. Bicknell[70], N. Biederbeck[46], C. Bigongiari[28], A. Biland[36], R. Bird[71], E. Bissaldi[72], J. Biteau[73], M. Bitossi[74], O. Blanch[41], M. Blank[50], J. Blazek[33], J. Bobin[75], C. Boccato[76], F. Bocchino[77], C. Boehm[78], M. Bohacova[33], C. Boisson[20], J. Boix[41], J.-P. Bolle[52], J. Bolmont[79], G. Bonanno[29], C. Bonavolontà[24], L. Bonneau Arbeletche[80], G. Bonnoli[12], P. Bordas[81], J. Borkowski[49], S. Bórquez[35], R. Bose[82], D. Bose[83], Z. Bosnjak[47], E. Bottacini[55], M. Böttcher[1], M.T. Botticella[84], C. Boutonnet[85], F. Bouyjou[75], V. Bozhilov[86], E. Bozzo[38], L. Brahimi[39], C. Braiding[43], S. Brau-Nogué[87], S. Breen[78], J. Bregeon[39], M. Breuhaus[53], A. Brill[9], W. Brisken[88], E. Brocato[28], A.M. Brown[5], K. Brügge[46], P. Brun[89], P. Brun[39], F. Brun[89], L. Brunetti[45], G. Brunetti[90], P. Bruno[29], A. Bruno[91], A. Bruzzese[6], N. Bucciantini[22], J. Buckley[82], R. Bühler[52], A. Bulgarelli[21], T. Bulik[92], M. Bünning[52], M. Bunse[46], M. Burton[93], A. Burtovoi[76], M. Buscemi[94], S. Buschjäger[46], G. Busetto[55], J. Buss[46], K. Byrum[26], A. Caccianiga[95], F. Cadoux[17], A. Calanducci[29], C. Calderón[3], J. Calvo Tovar[32], R. Cameron[96], P. Campaña[35], R. Canestrari[91], F. Cangemi[79], B. Cantlay[31], M. Capalbi[91], M. Capasso[9], M. Cappi[21], A. Caproni[97], R. Capuzzo-Dolcetta[28], P. Caraveo[61], V. Cárdenas[98], L. Cardiel[41], M. Cardillo[99], C. Carlile[100], S. Caroff[45], R. Carosi[74], A. Carosi[17], E. Carquín[35], M. Carrère[39], J.-M. Casandjian[4], S. Casanova[101,53], E. Cascone[84], F. Cassol[27], A.J. Castro-Tirado[12], F. Catalani[102], O. Catalano[91], D. Cauz[103], A. Ceccanti[64], C. Celestino Silva[80], S. Celli[18], K. Cerny[104], M. Cerruti[85], E. Chabanne[45], P. Chadwick[5], Y. Chai[105], P. Chambery[106], C. Champion[85], S. Chandra[1], S. Chaty[4], A. Chen[58], K. Cheng[2], M. Chernyakova[107], G. Chiaro[61], A. Chiavassa[64,108], M. Chikawa[2], V.R. Chitnis[109], J. Chudoba[33], L. Chytka[104], S. Cikota[47], A. Circiello[24,110], P. Clark[5], M. Çolak[41], E. Colombo[32], J. Colome[13], S. Colonges[85], A. Comastri[21], A. Compagnino[91], V. Conforti[21], E. Congiu[95], R. Coniglione[94], J. Conrad[111], F. Conte[53], J.L. Contreras[11], P. Coppi[112], R. Cornat[8], J. Coronado-Blazquez[14], J. Cortina[113], A. Costa[29], H. Costantini[27], G. Cotter[114], B. Courty[85], S. Covino[95], S. Crestan[61], P. Cristofari[20], R. Crocker[70], J. Croston[115], K. Cubuk[93], O. Cuevas[98], X. Cui[2], G. Cusumano[91], S. Cutini[23], A. D'Aì[91], G. D'Amico[116], F. D'Ammando[90], P. D'Avanzo[95], P. Da Vela[74], M. Dadina[21], S. Dai[117], M. Dalchenko[17], M. Dall' Ora[84], M.K. Daniel[63], J. Dauguet[85], I. Davids[48], J. Davies[114], B. Dawson[118], A. De Angelis[55], A.E. de Araújo Carvalho[40], M. de Bony de Lavergne[45], V. De Caprio[84], G. De Cesare[21], F. De Frondat[20], E.M. de Gouveia Dal Pino[19], I. de la Calle[11], B. De Lotto[103], A. De Luca[61], D. De Martino[84], R.M. de Menezes[19], M. de Naurois[8], E. de Oña Wilhelmi[13], F. De Palma[64], F. De Persio[119], N. de Simone[52], V. de Souza[80], M. Del Santo[91], M.V. del Valle[19], E. Delagnes[75], G. Deleglise[45], M. Delfino Reznicek[6], C. Delgado[113], A.G. Delgado Giler[80], J. Delgado Mengual[6], R. Della Ceca[95], M. Della Valle[84], D. della Volpe[17], D. Depaoli[64,108], D. Depouez[27], J. Devin[85], T. Di Girolamo[24,110], C. Di Giulio[25], A. Di Piano[21], F. Di Pierro[64], L. Di Venere[120], C. Díaz[113], C. Díaz-Bahamondes[3], C. Dib[35], S. Diebold[69], S. Digel[96], R. Dima[55], A. Djannati-Ataï[85], J. Djuvsland[116], A. Dmytriiev[20], K. Docher[9], A. Domínguez[11], D. Dominis Prester[121],

A. Donath[53], A. Donini[41], D. Dorner[122], M. Doro[55], R.d.C. dos Anjos[123], J.-L. Dournaux[20], T. Downes[107], G. Drake[26], H. Drass[3], D. Dravins[100], C. Duangchan[31], A. Duara[124], G. Dubus[125], L. Ducci[69], C. Duffy[124], D. Dumora[106], K. Dundas Morå[111], A. Durkalec[126], V.V. Dwarkadas[127], J. Ebr[33], C. Eckner[45], J. Eder[105], A. Ederoclite[19], E. Edy[8], K. Egberts[128], S. Einecke[118], J. Eisch[129], C. Eleftheriadis[130], D. Elsässer[46], G. Emery[17], D. Emmanoulopoulos[115], J.-P. Ernenwein[27], M. Errando[82], P. Escarate[35], J. Escudero[12], C. Espinoza[3], S. Ettori[21], A. Eungwanichayapant[31], P. Evans[124], C. Evoli[18], M. Fairbairn[131], D. Falceta-Goncalves[132], A. Falcone[133], V. Fallah Ramazani[65], R. Falomo[76], K. Farakos[134], G. Fasola[20], A. Fattorini[46], Y. Favre[17], R. Fedora[135], E. Fedorova[136], S. Fegan[8], K. Feijen[118], Q. Feng[9], G. Ferrand[54], G. Ferrara[94], O. Ferreira[8], M. Fesquet[75], E. Fiandrini[23], A. Fiasson[45], M. Filipovic[117], D. Fink[105], J.P. Finley[137], V. Fioretti[21], D.F.G. Fiorillo[24,110], M. Fiorini[61], S. Flis[52], H. Flores[20], L. Foffano[17], C. Föhr[53], M.V. Fonseca[11], L. Font[138], G. Fontaine[8], O. Fornieri[52], P. Fortin[63], L. Fortson[88], N. Fouque[45], A. Fournier[106], B. Fraga[40], A. Franceschini[76], F.J. Franco[30], A. Franco Ordovas[32], L. Freixas Coromina[113], L. Fresnillo[30], C. Fruck[105], D. Fugazza[95], Y. Fujikawa[139], Y. Fujita[2], S. Fukami[2], Y. Fukazawa[140], Y. Fukui[141], D. Fulla[52], S. Funk[142], A. Furniss[143], O. Gabella[39], S. Gabici[85], D. Gaggero[14], G. Galanti[61], G. Galaz[3], P. Galdemard[144], Y. Gallant[39], D. Galloway[7], S. Gallozzi[28], V. Gammaldi[14], R. Garcia[41], E. Garcia[45], E. García[13], R. Garcia López[32], M. Garczarczyk[52], F. Gargano[120], C. Gargano[91], S. Garozzo[29], D. Gascon[81], T. Gasparetto[145], D. Gasparrini[25], H. Gasparyan[52], M. Gaug[138], N. Geffroy[45], A. Gent[146], S. Germani[76], L. Gesa[13], A. Ghalumyan[147], A. Ghedina[148], G. Ghirlanda[95], F. Gianotti[21], S. Giarrusso[91], M. Giarrusso[94], G. Giavitto[52], B. Giebels[8], N. Giglietto[72], V. Gika[134], F. Gillardo[45], R. Gimenes[19], F. Giordano[149], G. Giovannini[90], E. Giro[76], M. Giroletti[90], A. Giuliani[61], L. Giunti[85], M. Gjaja[9], J.-F. Glicenstein[89], P. Gliwny[60], N. Godinovic[150], H. Göksu[53], P. Goldoni[85], J.L. Gómez[12], G. Gómez-Vargas[3], M.M. González[16], J.M. González[151], K.S. Gothe[109], D. Götz[4], J. Goulart Coelho[123], K. Gourgouliatos[5], T. Grabarczyk[152], R. Graciani[81], P. Grandi[21], G. Grasseau[8], D. Grasso[74], A.J. Green[78], D. Green[105], J. Green[28], T. Greenshaw[153], I. Grenier[4], P. Grespan[55], A. Grillo[29], M.-H. Grondin[106], J. Grube[131], V. Guarino[26], B. Guest[37], O. Gueta[52], M. Gündüz[59], S. Gunji[154], A. Gusdorf[20], G. Gyuk[155], J. Hackfeld[59], D. Hadasch[2], J. Haga[139], L. Hagge[52], A. Hahn[105], J.E. Hajlaoui[85], H. Hakobyan[35], A. Halim[89], P. Hamal[33], W. Hanlon[63], S. Hara[156], Y. Harada[157], M.J. Hardcastle[158], M. Harvey[5], K. Hashiyama[2], T. Hassan Collado[113], T. Haubold[105], A. Haupt[52], U.A. Hautmann[159], M. Havelka[33], K. Hayashi[141], K. Hayashi[160], M. Hayashida[161], H. He[54], L. Heckmann[105], M. Heller[17], J.C. Helo[35], F. Henault[125], G. Henri[125], G. Hermann[53], R. Hermel[45], S. Hernández Cadena[16], J. Herrera Llorente[32], A. Herrero[32], O. Hervet[143], J. Hinton[53], A. Hiramatsu[157], N. Hiroshima[54], K. Hirotani[2], B. Hnatyk[136], R. Hnatyk[136], J.K. Hoang[11], D. Hoffmann[27], W. Hofmann[53], C. Hoischen[128], J. Holder[162], M. Holler[163], B. Hona[164], D. Horan[8], J. Hörandel[165], D. Horns[50], P. Horvath[104], J. Houles[27], T. Hovatta[65], M. Hrabovsky[104], D. Hrupec[166], Y. Huang[135], J.-M. Huet[20], G. Hughes[159], D. Hui[2], G. Hull[73], T.B. Humensky[9], M. Hütten[105], R. Iaria[77], M. Iarlori[18], J.M. Illa[41], R. Imazawa[140], D. Impiombato[91], T. Inada[2], F. Incardona[29], A. Ingallinera[29], Y. Inome[2], S. Inoue[54], T. Inoue[141], Y. Inoue[167], A. Insolia[120,94], F. Iocco[24,110], K. Ioka[168], M. Ionica[23], M. Iori[119], S. Iovenitti[95], A. Iriarte[16], K. Ishio[105], W. Ishizaki[168], Y. Iwamura[2], C. Jablonski[105], J. Jacquemier[45], M. Jacquemont[45], M. Jamrozy[169], P. Janecek[33], F. Jankowsky[170], A. Jardin-Blicq[31], C. Jarnot[87], P. Jean[87], I. Jiménez Martínez[113], W. Jin[171], L. Jocou[125], N. Jordana[172], M. Josselin[73], L. Jouvin[41], I. Jung-Richardt[142], F.J.P.A. Junqueira[19], C. Juramy-Gilles[79], J. Jurysek[38], P. Kaaret[173], L.H.S. Kadowaki[19], M. Kagaya[2], O. Kalekin[142], R. Kankanyan[53], D. Kantzas[174], V. Karas[34], A. Karastergiou[114], S. Karkar[79], E. Kasai[48], J. Kasperek[175], H. Katagiri[176], J. Kataoka[177], K. Katarzyński[178], S. Katsuda[179], U. Katz[142], N. Kawanaka[180], D. Kazanas[130], D. Kerszberg[41], B. Khélifi[85], M.C. Kherlakian[52], T.P. Kian[181], D.B. Kieda[164], T. Kihm[53], S. Kim[3], S. Kimeswenger[163], S. Kisaka[140], R. Kissmann[163], R. Kleijwegt[135], T. Kleiner[52], G. Kluge[10], W. Kluźniak[49], J. Knapp[52], J. Knödlseder[87], A. Kobakhidze[78], Y. Kobayashi[2], B. Koch[3], J. Kocot[152], K. Kohri[182], K. Kokkotas[69], N. Komin[58], A. Kong[2], K. Kosack[4], G. Kowal[132], F. Krack[52], M. Krause[52], F. Krennrich[129], M. Krumholz[70], H. Kubo[180], V. Kudryavtsev[183], S. Kunwar[53], Y. Kuroda[139], J. Kushida[157], P. Kushwaha[19], A. La Barbera[91], N. La Palombara[61], V. La Parola[91], G. La Rosa[91], R. Lahmann[142], G. Lamanna[45], A. Lamastra[28],

M. Landoni[95], D. Landriu[4], R.G. Lang[80], J. Lapington[124], P. Laporte[20], P. Lason[152], J. Lasuik[37], J. Lazendic-Galloway[7], T. Le Flour[45], P. Le Sidaner[20], S. Leach[124], A. Leckngam[31], S.-H. Lee[180], W.H. Lee[16], S. Lee[118], M.A. Leigui de Oliveira[184], A. Lemière[85], M. Lemoine-Goumard[106], J.-P. Lenain[79], F. Leone[94,185], V. Leray[8], G. Leto[29], F. Leuschner[69], C. Levy[79,20], R. Lindemann[52], E. Lindfors[65], L. Linhoff[46], I. Liodakis[65], A. Lipniacka[116], S. Lloyd[5], M. Lobo[113], T. Lohse[186], S. Lombardi[28], F. Longo[145], A. Lopez[32], M. López[11], R. López-Coto[55], S. Loporchio[149], F. Louis[75], M. Louys[20], F. Lucarelli[28], D. Lucchesi[55], H. Ludwig Boudi[39], P.L. Luque-Escamilla[56], E. Lyard[38], M.C. Maccarone[91], T. Maccarone[187], E. Mach[101], A.J. Maciejewski[188], J. Mackey[15], G.M. Madejski[96], P. Maeght[39], C. Maggio[138], G. Maier[52], A. Majczyna[126], P. Majumdar[83,2], M. Makariev[189], M. Mallamaci[55], R. Malta Nunes de Almeida[184], S. Maltezos[134], D. Malyshev[142], D. Malyshev[69], D. Mandat[33], G. Maneva[189], M. Manganaro[121], G. Manicò[94], P. Manigot[8], K. Mannheim[122], N. Maragos[134], D. Marano[29], M. Marconi[84], A. Marcowith[39], M. Marculewicz[190], B. Marčun[68], J. Marín[98], N. Marinello[55], P. Marinos[118], M. Mariotti[55], S. Markoff[174], P. Marquez[41], G. Marsella[94], J. Martí[56], J.-M. Martin[20], P. Martin[87], O. Martinez[30], M. Martínez[41], G. Martínez[113], O. Martínez[41], H. Martínez-Huerta[80], C. Marty[87], R. Marx[53], N. Masetti[21,151], P. Massimino[29], A. Mastichiadis[191], H. Matsumoto[167], N. Matthews[164], G. Maurin[45], W. Max-Moerbeck[192], N. Maxted[43], D. Mazin[2,105], M.N. Mazziotta[120], S.M. Mazzola[77], J.D. Mbarubucyeye[52], L. Mc Comb[5], I. McHardy[115], S. McKeague[107], S. McMuldroch[63], E. Medina[64], D. Medina Miranda[17], A. Melandri[95], C. Melioli[19], D. Melkumyan[52], S. Menchiari[62], S. Mender[46], S. Mereghetti[61], G. Merino Arévalo[6], E. Mestre[13], J.-L. Meunier[79], T. Meures[135], M. Meyer[142], S. Micanovic[121], M. Miceli[77], M. Michailidis[69], J. Michałowski[101], T. Miener[11], I. Mievre[45], J. Miller[35], I.A. Minaya[153], T. Mineo[91], M. Minev[189], J.M. Miranda[30], R. Mirzoyan[105], A. Mitchell[36], T. Mizuno[193], B. Mode[135], R. Moderski[49], L. Mohrmann[142], E. Molina[81], E. Molinari[148], T. Montaruli[17], I. Monteiro[45], C. Moore[124], A. Moralejo[41], D. Morcuende-Parrilla[11], E. Moretti[41], L. Morganti[64], K. Mori[194], P. Moriarty[15], K. Morik[46], G. Morlino[22], P. Morris[114], A. Morselli[25], K. Mosshammer[52], P. Moya[192], R. Mukherjee[9], J. Muller[8], C. Mundell[172], J. Mundet[41], T. Murach[52], A. Muraczewski[49], H. Muraishi[195], K. Murase[2], I. Musella[84], A. Musumarra[120], A. Nagai[17], N. Nagar[196], S. Nagataki[54], T. Naito[156], T. Nakamori[154], K. Nakashima[142], K. Nakayama[51], N. Nakhjiri[13], G. Naletto[55], D. Naumann[52], L. Nava[95], R. Navarro[174], M.A. Nawaz[132], H. Ndiyavala[1], D. Neise[36], L. Nellen[16], R. Nemmen[19], M. Newbold[164], N. Neyroud[45], K. Ngernphat[31], T. Nguyen Trung[73], L. Nicastro[21], L. Nickel[46], J. Niemiec[101], D. Nieto[11], M. Nievas[32], C. Nigro[41], M. Nikołajuk[190], D. Ninci[41], K. Nishijima[157], K. Noda[2], Y. Nogami[176], S. Nolan[5], R. Nomura[2], R. Norris[117], D. Nosek[197], M. Nöthe[46], B. Novosyadlyj[198], V. Novotny[197], S. Nozaki[180], F. Nunio[144], P. O'Brien[124], K. Obara[176], R. Oger[85], Y. Ohira[51], M. Ohishi[2], S. Ohm[52], Y. Ohtani[2], T. Oka[180], N. Okazaki[2], A. Okumura[139,199], J.-F. Olive[87], C. Oliver[30], G. Olivera[52], B. Olmi[22], R.A. Ong[71], M. Orienti[90], R. Orito[200], M. Orlandini[21], S. Orlando[77], E. Orlando[145], J.P. Osborne[124], M. Ostrowski[169], N. Otte[146], E. Ovcharov[86], E. Owen[2], I. Oya[159], A. Ozieblo[152], M. Padovani[22], I. Pagano[29], A. Pagliaro[91], A. Paizis[61], M. Palatiello[145], M. Palatka[33], E. Palazzi[21], J.-L. Panazol[45], D. Paneque[105], B. Panes[3], S. Panny[163], F.R. Pantaleo[72], M. Panter[53], R. Paoletti[62], M. Paolillo[24,110], A. Papitto[28], A. Paravac[122], J.M. Paredes[81], G. Pareschi[95], N. Park[127], N. Parmiggiani[21], R.D. Parsons[186], P. Paśko[201], S. Patel[52], B. Patricelli[28], G. Pauletta[103], L. Pavletić[121], S. Pavy[8], A. Pe'er[105], M. Pech[33], M. Pecimotika[121], M.G. Pellegriti[120], P. Peñil Del Campo[11], M. Penno[52], A. Pepato[55], S. Perard[106], C. Perennes[55], G. Peres[77], M. Peresano[4], A. Pérez-Aguilera[11], J. Pérez-Romero[14], M.A. Pérez-Torres[12], M. Perri[28], M. Persic[103], S. Petrera[18], P.-O. Petrucci[125], O. Petruk[66], B. Peyaud[89], K. Pfrang[52], E. Pian[21], G. Piano[99], P. Piatteli[94], E. Pietropaolo[18], R. Pillera[149], B. Pilszyk[101], D. Pimentel[202], F. Pintore[91], C. Pio García[41], G. Pirola[64], F. Piron[39], A. Pisarski[190], S. Pita[85], M. Pohl[128], V. Poireau[45], P. Poledrelli[159], A. Pollo[126], M. Polo[113], C. Pongkitivanichkul[31], J. Porthault[144], J. Powell[171], D. Pozo[98], R.R. Prado[52], E. Prandini[55], P. Prasit[31], J. Prast[45], K. Pressard[73], G. Principe[90], C. Priyadarshi[41], N. Produit[38], D. Prokhorov[174], H. Prokoph[52], M. Prouza[33], H. Przybilski[101], E. Pueschel[52], G. Pühlhofer[69], I. Puljak[150], M.L. Pumo[94], M. Punch[85,57], F. Queiroz[203], J. Quinn[204], A. Quirrenbach[170], S. Rainò[149], P.J. Rajda[175], R. Rando[55], S. Razzaque[205], E. Rebert[20], S. Recchia[85], P. Reichherzer[59], O. Reimer[163], A. Reimer[163], A. Reisenegger[3,206], Q. Remy[53],

M. Renaud[39], T. Reposeur[106], B. Reville[53], J.-M. Reymond[75], J. Reynolds[15], W. Rhode[46], D. Ribeiro[9], M. Ribó[81], G. Richards[162], T. Richtler[196], J. Rico[41], F. Rieger[53], L. Riitano[135], V. Ripepi[84], M. Riquelme[192], D. Riquelme[35], S. Rivoire[39], V. Rizi[18], E. Roache[63], B. Röben[159], M. Roche[106], J. Rodriguez[4], G. Rodriguez Fernandez[25], J.C. Rodriguez Ramirez[19], J.J. Rodríguez Vázquez[113], F. Roepke[170], G. Rojas[207], L. Romanato[55], P. Romano[95], G. Romeo[29], F. Romero Lobato[11], C. Romoli[53], M. Roncadelli[103], S. Ronda[30], J. Rosado[11], A. Rosales de Leon[5], G. Rowell[118], B. Rudak[49], A. Rugliancich[74], J.E. Ruíz del Mazo[12], W. Rujopakarn[31], C. Rulten[5], C. Russell[3], F. Russo[21], I. Sadeh[52], E. Sæther Hatlen[10], S. Safi-Harb[37], L. Saha[11], P. Saha[208], V. Sahakian[147], S. Sailer[53], T. Saito[2], N. Sakaki[54], S. Sakurai[2], F. Salesa Greus[101], G. Salina[25], H. Salzmann[69], D. Sanchez[45], M. Sánchez-Conde[14], H. Sandaker[10], A. Sandoval[16], P. Sangiorgi[91], M. Sanguillon[39], H. Sano[2], M. Santander[171], A. Santangelo[69], E.M. Santos[202], R. Santos-Lima[19], A. Sanuy[81], L. Sapozhnikov[96], T. Saric[150], S. Sarkar[114], H. Sasaki[157], N. Sasaki[179], K. Satalecka[52], Y. Sato[209], F.G. Saturni[28], M. Sawada[54], U. Sawangwit[31], J. Schaefer[142], A. Scherer[3], J. Scherpenberg[105], P. Schipani[84], B. Schleicher[122], J. Schmoll[5], M. Schneider[143], H. Schoorlemmer[53], P. Schovanek[33], F. Schussler[89], B. Schwab[142], U. Schwanke[186], J. Schwarz[95], T. Schweizer[105], E. Sciacca[29], S. Scuderi[61], M. Seglar Arroyo[45], A. Segreto[91], I. Seitenzahl[43], D. Semikoz[85], O. Sergijenko[136], J.E. Serna Franco[16], M. Servillat[20], K. Seweryn[201], V. Sguera[21], A. Shalchi[37], R.Y. Shang[71], P. Sharma[73], R.C. Shellard[40], L. Sidoli[61], J. Sieiro[81], H. Siejkowski[152], J. Silk[114], A. Sillanpää[65], B.B. Singh[109], K.K. Singh[210], A. Sinha[39], C. Siqueira[80], G. Sironi[95], J. Sitarek[60], P. Sizun[75], V. Sliusar[38], A. Slowikowska[178], D. Sobczyńska[60], R.W. Sobrinho[184], H. Sol[20], G. Sottile[91], H. Spackman[114], A. Specovius[142], S. Spencer[114], G. Spengler[186], D. Spiga[95], A. Spolon[55], W. Springer[164], A. Stamerra[28], S. Stanič[68], R. Starling[124], Ł. Stawarz[169], R. Steenkamp[48], S. Stefanik[197], C. Stegmann[128], A. Steiner[52], S. Steinmassl[53], C. Stella[103], C. Steppa[128], R. Sternberger[52], M. Sterzel[152], C. Stevens[135], B. Stevenson[71], T. Stolarczyk[4], G. Stratta[21], U. Straumann[208], J. Strišković[166], M. Strzys[2], R. Stuik[174], M. Suchenek[211], Y. Suda[140], Y. Sunada[179], T. Suomijarvi[73], T. Suric[212], P. Sutcliffe[153], H. Suzuki[213], P. Świerk[101], T. Szepieniec[152], A. Tacchini[21], K. Tachihara[141], G. Tagliaferri[95], H. Tajima[139], N. Tajima[2], D. Tak[52], K. Takahashi[214], H. Takahashi[140], M. Takahashi[2], M. Takahashi[2], J. Takata[2], R. Takeishi[2], T. Tam[2], M. Tanaka[182], T. Tanaka[213], S. Tanaka[209], D. Tateishi[179], M. Tavani[99], F. Tavecchio[95], T. Tavernier[89], L. Taylor[135], A. Taylor[52], L.A. Tejedor[11], P. Temnikov[189], Y. Terada[179], K. Terauchi[180], J.C. Terrazas[192], R. Terrier[85], T. Terzic[121], M. Teshima[105,2], V. Testa[28], D. Thibaut[85], F. Thocquenne[75], W. Tian[2], L. Tibaldo[87], A. Tiengo[215], D. Tiziani[142], M. Tluczykont[50], C.J. Todero Peixoto[102], F. Tokanai[154], K. Toma[160], L. Tomankova[142], J. Tomastik[104], D. Tonev[189], M. Tornikoski[216], D.F. Torres[13], E. Torresi[21], G. Tosti[95], L. Tosti[23], T. Totani[51], N. Tothill[117], F. Toussenel[79], G. Tovmassian[16], P. Travnicek[33], C. Trichard[8], M. Trifoglio[21], A. Trois[95], S. Truzzi[62], A. Tsiahina[87], T. Tsuru[180], B. Turk[45], A. Tutone[91], Y. Uchiyama[161], G. Umana[29], P. Utayarat[31], L. Vaclavek[104], M. Vacula[104], V. Vagelli[23,217], F. Vagnetti[25], F. Vakili[218], J.A. Valdivia[192], M. Valentino[24], A. Valio[19], B. Vallage[89], P. Vallania[44,64], J.V. Valverde Quispe[8], A.M. Van den Berg[42], W. van Driel[20], C. van Eldik[142], C. van Rensburg[1], B. van Soelen[210], J. Vandenbroucke[135], J. Vanderwalt[1], G. Vasileiadis[39], V. Vassiliev[71], M. Vázquez Acosta[32], M. Vecchi[42], A. Vega[98], J. Veh[142], P. Veitch[118], P. Venault[75], C. Venter[1], S. Ventura[62], S. Vercellone[95], S. Vergani[20], V. Verguilov[189], G. Verna[27], S. Vernetto[44,64], V. Verzi[25], G.P. Vettolani[90], C. Veyssiere[144], I. Viale[55], A. Viana[80], N. Viaux[35], J. Vicha[33], J. Vignatti[35], C.F. Vigorito[64,108], J. Villanueva[98], J. Vink[174], V. Vitale[23], V. Vittorini[99], V. Vodeb[68], H. Voelk[53], N. Vogel[142], V. Voisin[79], S. Vorobiov[68], I. Vovk[2], M. Vrastil[33], T. Vuillaume[45], S.J. Wagner[170], R. Wagner[105], P. Wagner[52], K. Wakazono[139], S.P. Wakely[127], R. Walter[38], M. Ward[5], D. Warren[54], J. Watson[52], N. Webb[87], M. Wechakama[31], P. Wegner[52], A. Weinstein[129], C. Weniger[174], F. Werner[53], H. Wetteskind[105], M. White[118], R. White[53], A. Wierzcholska[101], S. Wiesand[52], R. Wijers[174], M. Wilkinson[124], M. Will[105], D.A. Williams[143], J. Williams[124], T. Williamson[162], A. Wolter[95], Y.W. Wong[142], M. Wood[96], C. Wunderlich[62], T. Yamamoto[213], H. Yamamoto[141], Y. Yamane[141], R. Yamazaki[209], S. Yanagita[176], L. Yang[205], S. Yoo[180], T. Yoshida[176], T. Yoshikoshi[2], P. Yu[71], P. Yu[85], A. Yusafzai[59], M. Zacharias[20], G. Zaharijas[68], B. Zaldivar[14], L. Zampieri[76], R. Zanmar Sanchez[29], D. Zaric[150], M. Zavrtanik[68], D. Zavrtanik[68], A.A. Zdziarski[49], A. Zech[20], H. Zechlin[64], A. Zenin[139], A. Zerwekh[35], V.I. Zhdanov[136],

K. Ziętara[169], A. Zink[142], J. Ziółkowski[49], V. Zitelli[21], M. Živec[68], A. Zmija[142]
1 : Centre for Space Research, North-West University, Potchefstroom, 2520, South Africa
2 : Institute for Cosmic Ray Research, University of Tokyo, 5-1-5, Kashiwa-no-ha, Kashiwa, Chiba 277-8582, Japan
3 : Pontificia Universidad Católica de Chile, Av. Libertador Bernardo O'Higgins 340, Santiago, Chile
4 : AIM, CEA, CNRS, Université Paris-Saclay, Université Paris Diderot, Sorbonne Paris Cité, CEA Paris-Saclay, IRFU/DAp, Bat 709, Orme des Merisiers, 91191 Gif-sur-Yvette, France
5 : Centre for Advanced Instrumentation, Dept. of Physics, Durham University, South Road, Durham DH1 3LE, United Kingdom
6 : Port d'Informació Científica, Edifici D, Carrer de l'Albareda, 08193 Bellaterrra (Cerdanyola del Vallès), Spain
7 : School of Physics and Astronomy, Monash University, Melbourne, Victoria 3800, Australia
8 : Laboratoire Leprince-Ringuet, École Polytechnique (UMR 7638, CNRS/IN2P3, Institut Polytechnique de Paris), 91128 Palaiseau, France
9 : Department of Physics, Columbia University, 538 West 120th Street, New York, NY 10027, USA
10 : University of Oslo, Department of Physics, Sem Saelandsvei 24 - PO Box 1048 Blindern, N-0316 Oslo, Norway
11 : EMFTEL department and IPARCOS, Universidad Complutense de Madrid, 28040 Madrid, Spain
12 : Instituto de Astrofísica de Andalucía-CSIC, Glorieta de la Astronomía s/n, 18008, Granada, Spain
13 : Institute of Space Sciences (ICE-CSIC), and Institut d'Estudis Espacials de Catalunya (IEEC), and Institució Catalana de Recerca I Estudis Avançats (ICREA), Campus UAB, Carrer de Can Magrans, s/n 08193 Cerdanyola del Vallés, Spain
14 : Instituto de Física Teórica UAM/CSIC and Departamento de Física Teórica, Universidad Autónoma de Madrid, c/ Nicolás Cabrera 13-15, Campus de Cantoblanco UAM, 28049 Madrid, Spain
15 : Dublin Institute for Advanced Studies, 31 Fitzwilliam Place, Dublin 2, Ireland
16 : Universidad Nacional Autónoma de México, Delegación Coyoacán, 04510 Ciudad de México, Mexico
17 : University of Geneva - Département de physique nucléaire et corpusculaire, 24 rue du Général-Dufour, 1211 Genève 4, Switzerland
18 : INFN Dipartimento di Scienze Fisiche e Chimiche - Università degli Studi dell'Aquila and Gran Sasso Science Institute, Via Vetoio 1, Viale Crispi 7, 67100 L'Aquila, Italy
19 : Instituto de Astronomia, Geofísico, e Ciências Atmosféricas - Universidade de São Paulo, Cidade Universitária, R. do Matão, 1226, CEP 05508-090, São Paulo, SP, Brazil
20 : LUTH, GEPI and LERMA, Observatoire de Paris, CNRS, PSL University, 5 place Jules Janssen, 92190, Meudon, France
21 : INAF - Osservatorio di Astrofisica e Scienza dello spazio di Bologna, Via Piero Gobetti 93/3, 40129 Bologna, Italy
22 : INAF - Osservatorio Astrofisico di Arcetri, Largo E. Fermi, 5 - 50125 Firenze, Italy
23 : INFN Sezione di Perugia and Università degli Studi di Perugia, Via A. Pascoli, 06123 Perugia, Italy
24 : INFN Sezione di Napoli, Via Cintia, ed. G, 80126 Napoli, Italy
25 : INFN Sezione di Roma Tor Vergata, Via della Ricerca Scientifica 1, 00133 Rome, Italy
26 : Argonne National Laboratory, 9700 S. Cass Avenue, Argonne, IL 60439, USA
27 : Aix-Marseille Université, CNRS/IN2P3, CPPM, 163 Avenue de Luminy, 13288 Marseille cedex 09, France
28 : INAF - Osservatorio Astronomico di Roma, Via di Frascati 33, 00040, Monteporzio Catone, Italy
29 : INAF - Osservatorio Astrofisico di Catania, Via S. Sofia, 78, 95123 Catania, Italy
30 : Grupo de Electronica, Universidad Complutense de Madrid, Av. Complutense s/n, 28040 Madrid, Spain

31 : National Astronomical Research Institute of Thailand, 191 Huay Kaew Rd., Suthep, Muang, Chiang Mai, 50200, Thailand
32 : Instituto de Astrofísica de Canarias and Departamento de Astrofísica, Universidad de La Laguna, La Laguna, Tenerife, Spain
33 : FZU - Institute of Physics of the Czech Academy of Sciences, Na Slovance 1999/2, 182 21 Praha 8, Czech Republic
34 : Astronomical Institute of the Czech Academy of Sciences, Bocni II 1401 - 14100 Prague, Czech Republic
35 : CCTVal, Universidad Técnica Federico Santa María, Avenida España 1680, Valparaíso, Chile
36 : ETH Zurich, Institute for Particle Physics, Schafmattstr. 20, CH-8093 Zurich, Switzerland
37 : The University of Manitoba, Dept of Physics and Astronomy, Winnipeg, Manitoba R3T 2N2, Canada
38 : Department of Astronomy, University of Geneva, Chemin d'Ecogia 16, CH-1290 Versoix, Switzerland
39 : Laboratoire Univers et Particules de Montpellier, Université de Montpellier, CNRS/IN2P3, CC 72, Place Eugène Bataillon, F-34095 Montpellier Cedex 5, France
40 : Centro Brasileiro de Pesquisas Físicas, Rua Xavier Sigaud 150, RJ 22290-180, Rio de Janeiro, Brazil
41 : Institut de Fisica d'Altes Energies (IFAE), The Barcelona Institute of Science and Technology, Campus UAB, 08193 Bellaterra (Barcelona), Spain
42 : University of Groningen, KVI - Center for Advanced Radiation Technology, Zernikelaan 25, 9747 AA Groningen, The Netherlands
43 : School of Physics, University of New South Wales, Sydney NSW 2052, Australia
44 : INAF - Osservatorio Astrofisico di Torino, Strada Osservatorio 20, 10025 Pino Torinese (TO), Italy
45 : Univ. Savoie Mont Blanc, CNRS, Laboratoire d'Annecy de Physique des Particules - IN2P3, 74000 Annecy, France
46 : Department of Physics, TU Dortmund University, Otto-Hahn-Str. 4, 44221 Dortmund, Germany
47 : University of Zagreb, Faculty of electrical engineering and computing, Unska 3, 10000 Zagreb, Croatia
48 : University of Namibia, Department of Physics, 340 Mandume Ndemufayo Ave., Pioneerspark, Windhoek, Namibia
49 : Nicolaus Copernicus Astronomical Center, Polish Academy of Sciences, ul. Bartycka 18, 00-716 Warsaw, Poland
50 : Universität Hamburg, Institut für Experimentalphysik, Luruper Chaussee 149, 22761 Hamburg, Germany
51 : Graduate School of Science, University of Tokyo, 7-3-1 Hongo, Bunkyo-ku, Tokyo 113-0033, Japan
52 : Deutsches Elektronen-Synchrotron, Platanenallee 6, 15738 Zeuthen, Germany
53 : Max-Planck-Institut für Kernphysik, Saupfercheckweg 1, 69117 Heidelberg, Germany
54 : RIKEN, Institute of Physical and Chemical Research, 2-1 Hirosawa, Wako, Saitama, 351-0198, Japan
55 : INFN Sezione di Padova and Università degli Studi di Padova, Via Marzolo 8, 35131 Padova, Italy
56 : Escuela Politécnica Superior de Jaén, Universidad de Jaén, Campus Las Lagunillas s/n, Edif. A3, 23071 Jaén, Spain
57 : Department of Physics and Electrical Engineering, Linnaeus University, 351 95 Växjö, Sweden
58 : University of the Witwatersrand, 1 Jan Smuts Avenue, Braamfontein, 2000 Johannesburg, South Africa
59 : Institut für Theoretische Physik, Lehrstuhl IV: Plasma-Astroteilchenphysik, Ruhr-Universität Bochum, Universitätsstraße 150, 44801 Bochum, Germany
60 : Faculty of Physics and Applied Computer Science, University of Lódź, ul. Pomorska 149-153, 90-236 Lódź, Poland
61 : INAF - Istituto di Astrofisica Spaziale e Fisica Cosmica di Milano, Via A. Corti 12, 20133 Milano, Italy
62 : INFN and Università degli Studi di Siena, Dipartimento di Scienze Fisiche, della Terra e dell'Ambiente (DSFTA), Sezione di Fisica, Via Roma 56, 53100 Siena, Italy
63 : Center for Astrophysics | Harvard & Smithsonian, 60 Garden St, Cambridge, MA 02180, USA

64 : INFN Sezione di Torino, Via P. Giuria 1, 10125 Torino, Italy
65 : Finnish Centre for Astronomy with ESO, University of Turku, Finland, FI-20014 University of Turku, Finland
66 : Pidstryhach Institute for Applied Problems in Mechanics and Mathematics NASU, 3B Naukova Street, Lviv, 79060, Ukraine
67 : Bhabha Atomic Research Centre, Trombay, Mumbai 400085, India
68 : Center for Astrophysics and Cosmology, University of Nova Gorica, Vipavska 11c, 5270 Ajdovščina, Slovenia
69 : Institut für Astronomie und Astrophysik, Universität Tübingen, Sand 1, 72076 Tübingen, Germany
70 : Research School of Astronomy and Astrophysics, Australian National University, Canberra ACT 0200, Australia
71 : Department of Physics and Astronomy, University of California, Los Angeles, CA 90095, USA
72 : INFN Sezione di Bari and Politecnico di Bari, via Orabona 4, 70124 Bari, Italy
73 : Laboratoire de Physique des 2 infinis, Irene Joliot-Curie,IN2P3/CNRS, Université Paris-Saclay, Université de Paris, 15 rue Georges Clemenceau, 91406 Orsay, Cedex, France
74 : INFN Sezione di Pisa, Largo Pontecorvo 3, 56217 Pisa, Italy
75 : IRFU/DEDIP, CEA, Université Paris-Saclay, Bat 141, 91191 Gif-sur-Yvette, France
76 : INAF - Osservatorio Astronomico di Padova, Vicolo dell'Osservatorio 5, 35122 Padova, Italy
77 : INAF - Osservatorio Astronomico di Palermo "G.S. Vaiana", Piazza del Parlamento 1, 90134 Palermo, Italy
78 : School of Physics, University of Sydney, Sydney NSW 2006, Australia
79 : Sorbonne Université, Université Paris Diderot, Sorbonne Paris Cité, CNRS/IN2P3, Laboratoire de Physique Nucléaire et de Hautes Energies, LPNHE, 4 Place Jussieu, F-75005 Paris, France
80 : Instituto de Física de São Carlos, Universidade de São Paulo, Av. Trabalhador São-carlense, 400 - CEP 13566-590, São Carlos, SP, Brazil
81 : Departament de Física Quàntica i Astrofísica, Institut de Ciències del Cosmos, Universitat de Barcelona, IEEC-UB, Martí i Franquès, 1, 08028, Barcelona, Spain
82 : Department of Physics, Washington University, St. Louis, MO 63130, USA
83 : Saha Institute of Nuclear Physics, Bidhannagar, Kolkata-700 064, India
84 : INAF - Osservatorio Astronomico di Capodimonte, Via Salita Moiariello 16, 80131 Napoli, Italy
85 : Université de Paris, CNRS, Astroparticule et Cosmologie, 10, rue Alice Domon et Léonie Duquet, 75013 Paris Cedex 13, France
86 : Astronomy Department of Faculty of Physics, Sofia University, 5 James Bourchier Str., 1164 Sofia, Bulgaria
87 : Institut de Recherche en Astrophysique et Planétologie, CNRS-INSU, Université Paul Sabatier, 9 avenue Colonel Roche, BP 44346, 31028 Toulouse Cedex 4, France
88 : School of Physics and Astronomy, University of Minnesota, 116 Church Street S.E. Minneapolis, Minnesota 55455-0112, USA
89 : IRFU, CEA, Université Paris-Saclay, Bât 141, 91191 Gif-sur-Yvette, France
90 : INAF - Istituto di Radioastronomia, Via Gobetti 101, 40129 Bologna, Italy
91 : INAF - Istituto di Astrofisica Spaziale e Fisica Cosmica di Palermo, Via U. La Malfa 153, 90146 Palermo, Italy
92 : Astronomical Observatory, Department of Physics, University of Warsaw, Aleje Ujazdowskie 4, 00478 Warsaw, Poland
93 : Armagh Observatory and Planetarium, College Hill, Armagh BT61 9DG, United Kingdom
94 : INFN Sezione di Catania, Via S. Sofia 64, 95123 Catania, Italy
95 : INAF - Osservatorio Astronomico di Brera, Via Brera 28, 20121 Milano, Italy

96 : Kavli Institute for Particle Astrophysics and Cosmology, Department of Physics and SLAC National Accelerator Laboratory, Stanford University, 2575 Sand Hill Road, Menlo Park, CA 94025, USA
97 : Universidade Cruzeiro do Sul, Núcleo de Astrofísica Teórica (NAT/UCS), Rua Galvão Bueno 8687, Bloco B, sala 16, Libertade 01506-000 - São Paulo, Brazil
98 : Universidad de Valparaíso, Blanco 951, Valparaiso, Chile
99 : INAF - Istituto di Astrofisica e Planetologia Spaziali (IAPS), Via del Fosso del Cavaliere 100, 00133 Roma, Italy
100 : Lund Observatory, Lund University, Box 43, SE-22100 Lund, Sweden
101 : The Henryk Niewodniczański Institute of Nuclear Physics, Polish Academy of Sciences, ul. Radzikowskiego 152, 31-342 Cracow, Poland
102 : Escola de Engenharia de Lorena, Universidade de São Paulo, Área I - Estrada Municipal do Campinho, s/n°, CEP 12602-810, Pte. Nova, Lorena, Brazil
103 : INFN Sezione di Trieste and Università degli Studi di Udine, Via delle Scienze 208, 33100 Udine, Italy
104 : Palacky University Olomouc, Faculty of Science, RCPTM, 17. listopadu 1192/12, 771 46 Olomouc, Czech Republic
105 : Max-Planck-Institut für Physik, Föhringer Ring 6, 80805 München, Germany
106 : CENBG, Univ. Bordeaux, CNRS-IN2P3, UMR 5797, 19 Chemin du Solarium, CS 10120, F-33175 Gradignan Cedex, France
107 : Dublin City University, Glasnevin, Dublin 9, Ireland
108 : Dipartimento di Fisica - Universitá degli Studi di Torino, Via Pietro Giuria 1 - 10125 Torino, Italy
109 : Tata Institute of Fundamental Research, Homi Bhabha Road, Colaba, Mumbai 400005, India
110 : Universitá degli Studi di Napoli "Federico II" - Dipartimento di Fisica "E. Pancini", Complesso universitario di Monte Sant'Angelo, Via Cintia - 80126 Napoli, Italy
111 : Oskar Klein Centre, Department of Physics, University of Stockholm, Albanova, SE-10691, Sweden
112 : Yale University, Department of Physics and Astronomy, 260 Whitney Avenue, New Haven, CT 06520-8101, USA
113 : CIEMAT, Avda. Complutense 40, 28040 Madrid, Spain
114 : University of Oxford, Department of Physics, Denys Wilkinson Building, Keble Road, Oxford OX1 3RH, United Kingdom
115 : School of Physics & Astronomy, University of Southampton, University Road, Southampton SO17 1BJ, United Kingdom
116 : Department of Physics and Technology, University of Bergen, Museplass 1, 5007 Bergen, Norway
117 : Western Sydney University, Locked Bag 1797, Penrith, NSW 2751, Australia
118 : School of Physical Sciences, University of Adelaide, Adelaide SA 5005, Australia
119 : INFN Sezione di Roma La Sapienza, P.le Aldo Moro, 2 - 00185 Roma, Italy
120 : INFN Sezione di Bari, via Orabona 4, 70126 Bari, Italy
121 : University of Rijeka, Department of Physics, Radmile Matejcic 2, 51000 Rijeka, Croatia
122 : Institute for Theoretical Physics and Astrophysics, Universität Würzburg, Campus Hubland Nord, Emil-Fischer-Str. 31, 97074 Würzburg, Germany
123 : Universidade Federal Do Paraná - Setor Palotina, Departamento de Engenharias e Exatas, Rua Pioneiro, 2153, Jardim Dallas, CEP: 85950-000 Palotina, Paraná, Brazil
124 : Dept. of Physics and Astronomy, University of Leicester, Leicester, LE1 7RH, United Kingdom
125 : Univ. Grenoble Alpes, CNRS, IPAG, 414 rue de la Piscine, Domaine Universitaire, 38041 Grenoble Cedex 9, France
126 : National Centre for nuclear research (Narodowe Centrum Badań Jądrowych), Ul. Andrzeja Sołtana7, 05-400 Otwock, Świerk, Poland

127 : Enrico Fermi Institute, University of Chicago, 5640 South Ellis Avenue, Chicago, IL 60637, USA
128 : Institut für Physik & Astronomie, Universität Potsdam, Karl-Liebknecht-Strasse 24/25, 14476 Potsdam, Germany
129 : Department of Physics and Astronomy, Iowa State University, Zaffarano Hall, Ames, IA 50011-3160, USA
130 : School of Physics, Aristotle University, Thessaloniki, 54124 Thessaloniki, Greece
131 : King's College London, Strand, London, WC2R 2LS, United Kingdom
132 : Escola de Artes, Ciências e Humanidades, Universidade de São Paulo, Rua Arlindo Bettio, CEP 03828-000, 1000 São Paulo, Brazil
133 : Dept. of Astronomy & Astrophysics, Pennsylvania State University, University Park, PA 16802, USA
134 : National Technical University of Athens, Department of Physics, Zografos 9, 15780 Athens, Greece
135 : University of Wisconsin, Madison, 500 Lincoln Drive, Madison, WI, 53706, USA
136 : Astronomical Observatory of Taras Shevchenko National University of Kyiv, 3 Observatorna Street, Kyiv, 04053, Ukraine
137 : Department of Physics, Purdue University, West Lafayette, IN 47907, USA
138 : Unitat de Física de les Radiacions, Departament de Física, and CERES-IEEC, Universitat Autònoma de Barcelona, Edifici C3, Campus UAB, 08193 Bellaterra, Spain
139 : Institute for Space-Earth Environmental Research, Nagoya University, Chikusa-ku, Nagoya 464-8601, Japan
140 : Department of Physical Science, Hiroshima University, Higashi-Hiroshima, Hiroshima 739-8526, Japan
141 : Department of Physics, Nagoya University, Chikusa-ku, Nagoya, 464-8602, Japan
142 : Friedrich-Alexander-Universität Erlangen-Nürnberg, Erlangen Centre for Astroparticle Physics (ECAP), Erwin-Rommel-Str. 1, 91058 Erlangen, Germany
143 : Santa Cruz Institute for Particle Physics and Department of Physics, University of California, Santa Cruz, 1156 High Street, Santa Cruz, CA 95064, USA
144 : IRFU / DIS, CEA, Université de Paris-Saclay, Bat 123, 91191 Gif-sur-Yvette, France
145 : INFN Sezione di Trieste and Università degli Studi di Trieste, Via Valerio 2 I, 34127 Trieste, Italy
146 : School of Physics & Center for Relativistic Astrophysics, Georgia Institute of Technology, 837 State Street, Atlanta, Georgia, 30332-0430, USA
147 : Alikhanyan National Science Laboratory, Yerevan Physics Institute, 2 Alikhanyan Brothers St., 0036, Yerevan, Armenia
148 : INAF - Telescopio Nazionale Galileo, Roche de los Muchachos Astronomical Observatory, 38787 Garafia, TF, Italy
149 : INFN Sezione di Bari and Università degli Studi di Bari, via Orabona 4, 70124 Bari, Italy
150 : University of Split - FESB, R. Boskovica 32, 21 000 Split, Croatia
151 : Universidad Andres Bello, República 252, Santiago, Chile
152 : Academic Computer Centre CYFRONET AGH, ul. Nawojki 11, 30-950 Cracow, Poland
153 : University of Liverpool, Oliver Lodge Laboratory, Liverpool L69 7ZE, United Kingdom
154 : Department of Physics, Yamagata University, Yamagata, Yamagata 990-8560, Japan
155 : Astronomy Department, Adler Planetarium and Astronomy Museum, Chicago, IL 60605, USA
156 : Faculty of Management Information, Yamanashi-Gakuin University, Kofu, Yamanashi 400-8575, Japan
157 : Department of Physics, Tokai University, 4-1-1, Kita-Kaname, Hiratsuka, Kanagawa 259-1292, Japan
158 : Centre for Astrophysics Research, Science & Technology Research Institute, University of Hertfordshire, College Lane, Hertfordshire AL10 9AB, United Kingdom
159 : Cherenkov Telescope Array Observatory, Saupfercheckweg 1, 69117 Heidelberg, Germany
160 : Tohoku University, Astronomical Institute, Aobaku, Sendai 980-8578, Japan

161 : Department of Physics, Rikkyo University, 3-34-1 Nishi-Ikebukuro, Toshima-ku, Tokyo, Japan
162 : Department of Physics and Astronomy and the Bartol Research Institute, University of Delaware, Newark, DE 19716, USA
163 : Institut für Astro- und Teilchenphysik, Leopold-Franzens-Universität, Technikerstr. 25/8, 6020 Innsbruck, Austria
164 : Department of Physics and Astronomy, University of Utah, Salt Lake City, UT 84112-0830, USA
165 : IMAPP, Radboud University Nijmegen, P.O. Box 9010, 6500 GL Nijmegen, The Netherlands
166 : Josip Juraj Strossmayer University of Osijek, Trg Ljudevita Gaja 6, 31000 Osijek, Croatia
167 : Department of Earth and Space Science, Graduate School of Science, Osaka University, Toyonaka 560-0043, Japan
168 : Yukawa Institute for Theoretical Physics, Kyoto University, Kyoto 606-8502, Japan
169 : Astronomical Observatory, Jagiellonian University, ul. Orla 171, 30-244 Cracow, Poland
170 : Landessternwarte, Zentrum für Astronomie der Universität Heidelberg, Königstuhl 12, 69117 Heidelberg, Germany
171 : University of Alabama, Tuscaloosa, Department of Physics and Astronomy, Gallalee Hall, Box 870324 Tuscaloosa, AL 35487-0324, USA
172 : Department of Physics, University of Bath, Claverton Down, Bath BA2 7AY, United Kingdom
173 : University of Iowa, Department of Physics and Astronomy, Van Allen Hall, Iowa City, IA 52242, USA
174 : Anton Pannekoek Institute/GRAPPA, University of Amsterdam, Science Park 904 1098 XH Amsterdam, The Netherlands
175 : Faculty of Computer Science, Electronics and Telecommunications, AGH University of Science and Technology, Kraków, al. Mickiewicza 30, 30-059 Cracow, Poland
176 : Faculty of Science, Ibaraki University, Mito, Ibaraki, 310-8512, Japan
177 : Faculty of Science and Engineering, Waseda University, Shinjuku, Tokyo 169-8555, Japan
178 : Institute of Astronomy, Faculty of Physics, Astronomy and Informatics, Nicolaus Copernicus University in Toruń, ul. Grudziądzka 5, 87-100 Toruń, Poland
179 : Graduate School of Science and Engineering, Saitama University, 255 Simo-Ohkubo, Sakura-ku, Saitama city, Saitama 338-8570, Japan
180 : Division of Physics and Astronomy, Graduate School of Science, Kyoto University, Sakyo-ku, Kyoto, 606-8502, Japan
181 : Centre for Quantum Technologies, National University Singapore, Block S15, 3 Science Drive 2, Singapore 117543, Singapore
182 : Institute of Particle and Nuclear Studies, KEK (High Energy Accelerator Research Organization), 1-1 Oho, Tsukuba, 305-0801, Japan
183 : Department of Physics and Astronomy, University of Sheffield, Hounsfield Road, Sheffield S3 7RH, United Kingdom
184 : Centro de Ciências Naturais e Humanas, Universidade Federal do ABC, Av. dos Estados, 5001, CEP: 09.210-580, Santo André - SP, Brazil
185 : Dipartimento di Fisica e Astronomia, Sezione Astrofisica, Universitá di Catania, Via S. Sofia 78, I-95123 Catania, Italy
186 : Department of Physics, Humboldt University Berlin, Newtonstr. 15, 12489 Berlin, Germany
187 : Texas Tech University, 2500 Broadway, Lubbock, Texas 79409-1035, USA
188 : University of Zielona Góra, ul. Licealna 9, 65-417 Zielona Góra, Poland
189 : Institute for Nuclear Research and Nuclear Energy, Bulgarian Academy of Sciences, 72 boul. Tsarigradsko chaussee, 1784 Sofia, Bulgaria
190 : University of Białystok, Faculty of Physics, ul. K. Ciołkowskiego 1L, 15-254 Białystok, Poland

191 : Faculty of Physics, National and Kapodestrian University of Athens, Panepistimiopolis, 15771 Ilissia, Athens, Greece
192 : Universidad de Chile, Av. Libertador Bernardo O'Higgins 1058, Santiago, Chile
193 : Hiroshima Astrophysical Science Center, Hiroshima University, Higashi-Hiroshima, Hiroshima 739-8526, Japan
194 : Department of Applied Physics, University of Miyazaki, 1-1 Gakuen Kibana-dai Nishi, Miyazaki, 889-2192, Japan
195 : School of Allied Health Sciences, Kitasato University, Sagamihara, Kanagawa 228-8555, Japan
196 : Departamento de Astronomía, Universidad de Concepción, Barrio Universitario S/N, Concepción, Chile
197 : Charles University, Institute of Particle & Nuclear Physics, V Holešovičkách 2, 180 00 Prague 8, Czech Republic
198 : Astronomical Observatory of Ivan Franko National University of Lviv, 8 Kyryla i Mephodia Street, Lviv, 79005, Ukraine
199 : Kobayashi-Maskawa Institute (KMI) for the Origin of Particles and the Universe, Nagoya University, Chikusa-ku, Nagoya 464-8602, Japan
200 : Graduate School of Technology, Industrial and Social Sciences, Tokushima University, Tokushima 770-8506, Japan
201 : Space Research Centre, Polish Academy of Sciences, ul. Bartycka 18A, 00-716 Warsaw, Poland
202 : Instituto de Física - Universidade de São Paulo, Rua do Matão Travessa R Nr.187 CEP 05508-090 Cidade Universitária, São Paulo, Brazil
203 : International Institute of Physics at the Federal University of Rio Grande do Norte, Campus Universitário, Lagoa Nova CEP 59078-970 Rio Grande do Norte, Brazil
204 : University College Dublin, Belfield, Dublin 4, Ireland
205 : Centre for Astro-Particle Physics (CAPP) and Department of Physics, University of Johannesburg, PO Box 524, Auckland Park 2006, South Africa
206 : Departamento de Física, Facultad de Ciencias Básicas, Universidad Metropolitana de Ciencias de la Educación, Santiago, Chile
207 : Núcleo de Formação de Professores - Universidade Federal de São Carlos, Rodovia Washington Luís, km 235 CEP 13565-905 - SP-310 São Carlos - São Paulo, Brazil
208 : Physik-Institut, Universität Zürich, Winterthurerstrasse 190, 8057 Zürich, Switzerland
209 : Department of Physical Sciences, Aoyama Gakuin University, Fuchinobe, Sagamihara, Kanagawa, 252-5258, Japan
210 : University of the Free State, Nelson Mandela Avenue, Bloemfontein, 9300, South Africa
211 : Faculty of Electronics and Information, Warsaw University of Technology, ul. Nowowiejska 15/19, 00-665 Warsaw, Poland
212 : Rudjer Boskovic Institute, Bijenicka 54, 10 000 Zagreb, Croatia
213 : Department of Physics, Konan University, Kobe, Hyogo, 658-8501, Japan
214 : Kumamoto University, 2-39-1 Kurokami, Kumamoto, 860-8555, Japan
215 : University School for Advanced Studies IUSS Pavia, Palazzo del Broletto, Piazza della Vittoria 15, 27100 Pavia, Italy
216 : Aalto University, Otakaari 1, 00076 Aalto, Finland
217 : Agenzia Spaziale Italiana (ASI), 00133 Roma, Italy
218 : Observatoire de la Cote d'Azur, Boulevard de l'Observatoire CS34229, 06304 Nice Cedex 4, Franc